\documentclass[twocolumn,longauthor]{aastex61}
\usepackage{amsmath}
\usepackage{units}

\newcommand{\E}[1]{\times10^{#1}}
\newcommand{\msol}{ \, M_\sun }
\newcommand{\mch}{ M_{\rm Ch} }
\newcommand{\dmb}{ \Delta {\rm m}_{15}(B) }
\newcommand{\bi}{\begin{itemize}}
\newcommand{\ei}{\end{itemize}}
\newcommand{\commentOut}[1]{}

\newcommand{\ia}{Ia\ }

\shorttitle{EVOLUTION OF THE SN \ia LUMINOSITY FUNCTION}
\shortauthors{SHEN, TOONEN, \& GRAUR}

\begin{document}

\title{The Evolution of the Type \ia Supernova Luminosity Function}

\author{Ken J. Shen}
\affiliation{Department of Astronomy and Theoretical Astrophysics Center, University of California, Berkeley, CA 94720, USA}

\author{Silvia Toonen}
\affiliation{Anton Pannekoek Institute for Astronomy, University of Amsterdam, 1090 GE Amsterdam, The Netherlands}
\affiliation{Department of Astronomy and Theoretical Astrophysics Center, University of California, Berkeley, CA 94720, USA}

\author{Or Graur}
\altaffiliation{NSF Astronomy and Astrophysics Postdoctoral Fellow.}
\affiliation{Harvard-Smithsonian Center for Astrophysics, 60 Garden St., Cambridge, MA 02138, USA}
\affiliation{Department of Astrophysics, American Museum of Natural History, New York, NY 10024, USA}

\correspondingauthor{Ken J. Shen}
\email{kenshen@astro.berkeley.edu}

\begin{abstract}

Type Ia supernovae (SNe~Ia) exhibit a wide diversity of peak luminosities and light curve shapes: the faintest SNe~Ia are $10$ times less luminous and evolve  more rapidly than the brightest SNe~Ia.  Their differing characteristics also extend to their stellar age distributions, with fainter SNe~Ia preferentially occurring in old stellar populations and vice versa.  In this Letter, we quantify this SN~Ia luminosity -- stellar age connection using data from the Lick Observatory Supernova Search (LOSS).  Our binary population synthesis calculations agree qualitatively with the observed trend in the $> 1 \,$Gyr-old populations probed by LOSS if the majority of SNe~Ia arise from prompt detonations of sub-Chandrasekhar mass white dwarfs (WDs) in double WD systems.  Under appropriate assumptions, we show that double WD systems with less massive primaries, which yield fainter SNe~Ia, interact and explode at older ages than those with more massive primaries.  We find that prompt detonations in double WD systems are capable of reproducing the observed evolution of the SN~Ia luminosity function, a constraint that any SN~Ia progenitor scenario must confront.

\end{abstract}


% -----------------------------------------------------------
% -----------------------------------------------------------

\section{Introduction}
\label{sec:intro}

Type Ia supernovae (SNe~Ia) are often referred to as ``standard candles.''  However, their intrinsic light curves vary significantly: bright SN 1991T-like SNe~Ia are $10$ times more luminous and evolve more slowly than the faint SN 1991bg-likes (see \citealt{taub17a} for a review).  The relationship between intrinsic luminosity and light curve shape is often referred to as the \cite{phil93a} relation, and it forms the basis for the use of SNe~Ia as cosmological distance indicators.

Brighter and fainter SNe~Ia also differ in their host galaxy distributions: bright SNe~Ia occur more often in low mass spiral galaxies, while faint SNe~Ia prefer high mass ellipticals \citep{hamu95a,sull06a,grau17b}.  While the range of progenitor metallicities may account for some of the dispersion in the Phillips relation, no amount of metallicity variation can account for the entire SN~Ia luminosity range for any progenitor scenario \citep{tbt03,shen17b}.  Thus, studies have suggested that the difference in host galaxy distributions of SN~Ia subtypes is due to the differing ages of the underlying stellar populations.

Linking stellar age to SN luminosity for Chandrasekhar-mass ($\mch$) explosion models has not been extensively studied (for one example, see \citealt{wang14a}) and appears difficult, if not impossible, to achieve.  Adjusting various quantities (e.g., the density at which the deflagration transitions to a detonation or the number of initial deflagration kernels) does not produce the relatively tight correlation of the Phillips relation and also fails to yield the low luminosity, rapidly evolving SN 1991bg-likes (\citealt{sim13a,blon17a}; although see \citealt{hoef17a}).  Since $\mch$ explosions do not reproduce the full range of the Phillips relation, connecting the stellar age to the various SN~Ia subtypes is as yet impossible within the $\mch$ paradigm.  Furthermore,  it is not  obvious why the deflagration-to-detonation transition density or number of ignition kernels would change with age.  Note that the category of $\mch$ explosion models includes both standard ``single degenerate'' scenarios (e.g., \citealt{wi73}) as well as ``double degenerate'' scenarios (e.g., \citealt{webb84}) for which the ignition occurs at the center of a super-$\mch$ merger remnant, as these have the same explosion mechanism and  similar radiative output.

At first glance, prospects appear better for sub-$\mch$ explosion models, in which the luminosity of the SN~Ia is directly related to the mass of the exploding WD \citep{sim10,blon17a,shen17b}, a quantity that could conceivably vary with stellar age.  Na\"{i}vely, it seems obvious that the masses of exploding sub-$\mch$ WDs decrease with age, because WD masses are directly related to main sequences masses, which are inversely related to main sequence lifetimes, and thus dimmer SNe~Ia would occur in older stellar populations as observed.

However, half of all SNe~Ia occur $> \unit [1]{Gyr}$ after their progenitor systems form (e.g., \citealt{maoz14a} and references therein), much longer than the main sequence lifetimes of the stars that produce the $ \gtrsim 0.85 \msol$ WDs that yield SNe~Ia.  For  sub-$\mch$ explosions produced by double WD binaries, either by double detonations \citep{guil10} or direct carbon ignitions \citep{pakm10},  the age of the system at the time of interaction is instead dominated by the gravitational wave inspiral timescale, which is itself a complicated outcome of multiple phases of stable and unstable mass transfer prior to the formation of the double WD system.   Note that sub-$\mch$ double detonation explosions may also occur in single degenerate systems in which the donor is a non-degenerate helium-rich star (e.g., \citealt{wtw86}) or in triple star systems \citep{kush13a}; however, because predicted rates from these systems are much lower than the SN~Ia rate \citep{geie13a,toon17b}, we restrict ourselves throughout the rest of this work to sub-$\mch$ explosions in isolated double WD systems.

In this Letter, for the first time, we quantify the evolution of exploding WD masses and resulting SN~Ia subtypes for sub-$\mch$ double WD progenitors and compare to observational constraints.\footnote{We note that \cite{ruit13a} and \cite{piro14b} also studied the SN~Ia luminosity function but did not analyze its evolution with time.}  In \S \ref{sec:loss}, we describe our basis for comparison: SN~Ia subtypes and stellar age distributions inferred from the Lick Observatory Supernova Search (LOSS) survey.  In \S \ref{sec:seba}, we detail the methodology by which we derive the theoretical SN~Ia subtype evolution from the \texttt{SeBa} binary population synthesis code.  We conclude and outline future work in \S \ref{sec:conc}.

% -----------------------------------------------------------
% -----------------------------------------------------------

\section{Observed evolution of the luminosity function}
\label{sec:loss}

During its first decade of operations, LOSS discovered more than 1000 SNe in the 14,882 galaxies it surveyed (e.g., \citealt{leam11a,li11b}).  \cite{li11b} constructed a volume-limited subsample that included 180 SNe and SN impostors. All SNe were classified spectroscopically, and individual SN light curves were used to calculate completeness corrections. The resulting sample is complete for SNe~Ia out to 80 Mpc. The SNe in this volume-limited sample were recently reclassified, based on additional data and an updated understanding of SN physics, but SNe~Ia were unaffected \citep{grau17a,grau17b,shiv17a}.

The LOSS volume-limited sample is homogeneous, well-characterized, and spectroscopically complete. However, LOSS targeted massive, luminous galaxies, so that low-luminosity galaxies and SN 1991T-like SNe~Ia, which are known to preferentially occur in these galaxies, are underrepresented.  With this in mind, we restrict our comparisons to the galactic ages $ > \unit[1]{Gyr}$ that are well-sampled in  LOSS.  Future work will use data from volume-limited samples that include more SNe~Ia in low-luminosity galaxies, which will allow us to better probe the early evolution of the luminosity function.

Of the 74 SNe~Ia in the updated volume-limited sample, we use the 70 SNe~Ia that were classified as ``normal,'' SN 1991bg-like, SN 1991T-like, or SN 1999aa-like. We exclude SNe 1999bh, 2002es, 2005cc, and 2005hk, which were classified as either SN 2002es-like or SN 2002cx-like.

Instead of relying on the discrete spectroscopic classifications of the SNe, we use the continuous  and extinction-independent  scale afforded by the $ \dmb $ parameter, which measures the decrease in $B$-band magnitudes between peak and $ \unit[15]{d}$ after peak. Through the \citet{phil93a} width-luminosity relation, this parameter is a good proxy for the intrinsic luminosity of a SN~Ia. Fifty-four SNe have $ \dmb $ measurements performed by different groups \citep{hick09a,cont10a,gane13a}. Twenty-six SNe did not have enough points on their light curves to fit for $  \dmb $ (J.~M. Silverman and W. Zhang, private communication). To fill in these missing values, we perform a linear fit between the extant $ \dmb $ values and the light-curve template number assigned to each LOSS SN by \citet{li11b}.

Next, we estimate the ages of the SN host galaxies by making use of the correlation between a galaxy's age and its stellar mass (e.g., \citealt{gall08a}).  We acknowledge that this relationship has large variance and that, furthermore, the average galaxy age is at best a rough proxy for the SN~Ia progenitor's age.  We leave a more accurate derivation of SN~Ia progenitor age to future work.

LOSS estimated host-galaxy stellar masses based on their $B$- and $K$-band luminosities \citep{leam11a}, but four of our host galaxies lack such estimates; they are assigned stellar masses using the method outlined by \cite{grau17b}.  These masses are then used to estimate stellar ages using Sloan Digital Sky Survey (SDSS) data (\citealt{york00}; \citealt{gall08a} and private communication; \citealt{calu14a}).

We can further refine our stellar age estimates by also using the morphological information of the galaxies.  \cite{gonz15a} present luminosity-weighted ages for a range of galaxy masses and Hubble types using data from the Calar Alto Legacy Integral Field Area (CALIFA) survey.  We interpolate among their results and apply a constant $+0.35$ dex correction to convert from luminosity- to mass-weighted ages \citep{godd17a}, which are more appropriate for the $> \unit[1]{Gyr}$ progenitors we consider.  In the following section, we compare theoretical CDFs of SN~Ia luminosities to observed CDFs for binned ages inferred from both methods.

% -----------------------------------------------------------
% -----------------------------------------------------------

\section{Theoretical evolution of the luminosity function}
\label{sec:seba}

In order to predict the evolution of SN~Ia subtypes from binary population synthesis calculations, we must construct a mapping from exploding WD mass, $M_1$, to $\dmb $, our observational proxy.  Radiative transfer simulations of a suite of sub-$\mch$ explosions were first performed by \cite{sim10}.  Recently, \citet[hereafter, S17]{shen17b} reexamined the subject using more precise detonation calculations and found significant differences in the nucleosynthetic products.  In complementary work, \citet[hereafter, B17]{blon17a} used a simplified nuclear network but improved upon the radiative transfer by employing a non-local thermodynamic equilibrium (non-LTE) code; they also found significant differences compared to \cite{sim10}.

None of the aforementioned studies was able to completely reproduce the Phillips relation: \cite{sim10} and S17 derived light curves confined to high values of $\dmb $, and while B17 found a good match to the Phillips relation in the high luminosity, low $\dmb $ regime, they were unable to achieve the high values of $\dmb $ at faint luminosities.  However, there are good reasons to believe that a combination of S17's nucleosynthesis and a non-LTE radiative transfer calculation like B17's will reproduce the Phillips relation.  S17's more detailed nucleosynthesis does not differ too substantially from that of B17 for higher WD masses $\simeq 1.1 \msol$, so a combination of the two improvements will not significantly alter B17's good agreement with observations of bright SNe~Ia.  At lower WD masses $ \leq 0.9 \msol$, S17's nucleosynthesis produces $\sim 3$ times more $^{56}$Ni than B17's.  Thus, a similar amount of $^{56}$Ni is produced in an explosion with a smaller ejecta mass, which implies a more rapid light curve evolution and higher values of $ \dmb$ at low luminosities, pushing B17's non-LTE calculations in the right direction.

\begin{figure}
  \centering
  \includegraphics[width=\columnwidth]{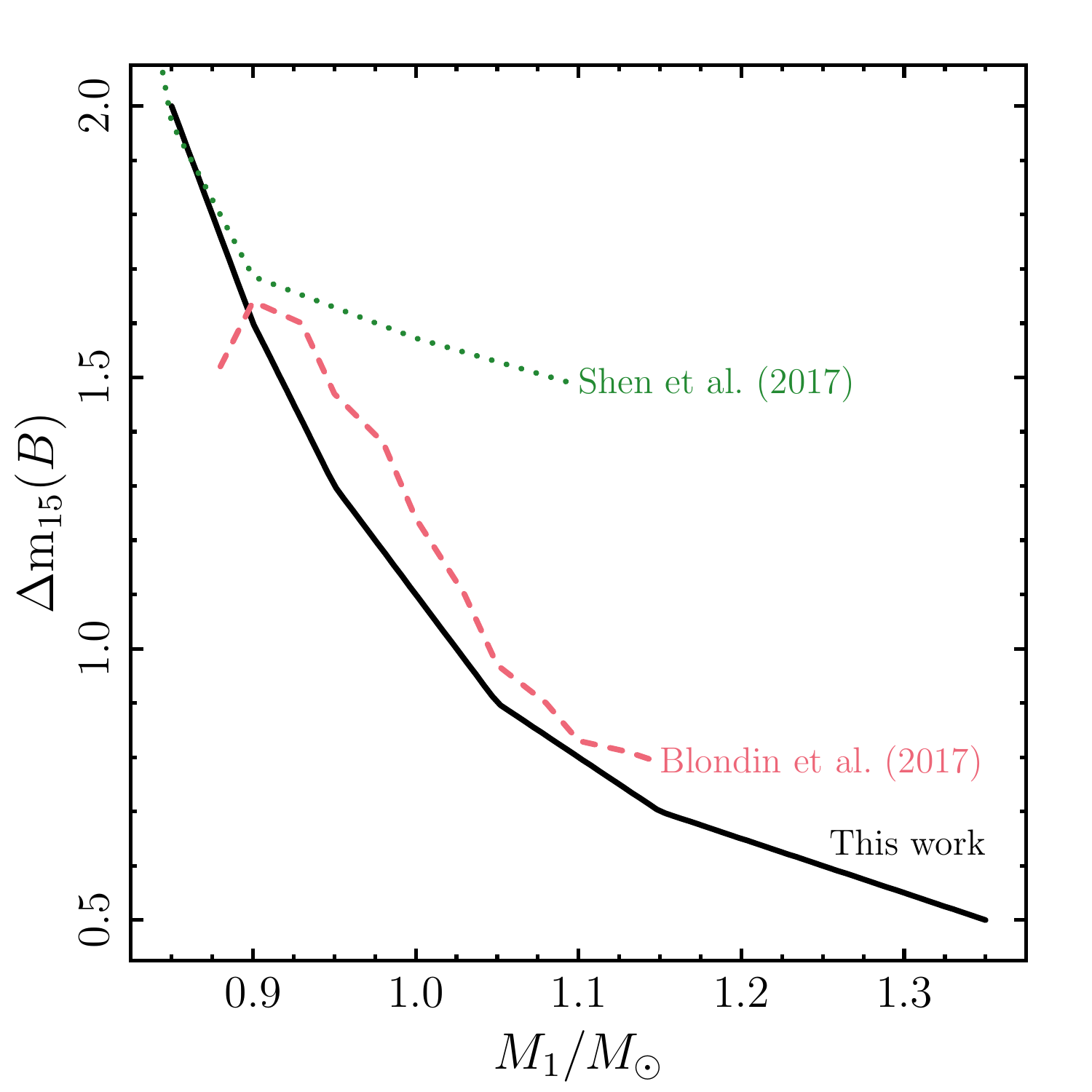}
  \caption{Assumed mapping of $M_1$ to $\dmb $ (\emph{solid line}).  A combination of the results from \cite{shen17b} (\emph{dotted line}) and \cite{blon17a} (\emph{dashed line}) is used to infer the mapping.}
  \label{fig:dm15map}
\end{figure}

Confirmation of the ability of sub-$\mch$ explosions to reproduce the entirety of the  Phillips relation awaits future calculations combining detailed nucleosynthesis with non-LTE radiative transfer.  For the remainder of this work, we assume that this effort will be successful and construct an appropriate mapping of exploding WD mass to $ \dmb$.  We assume SN 1991bg-likes with $\dmb = \unit[2.0]{mag} $ are produced by the explosions of $0.85 \msol$ WDs, as found by S17.  At the opposite end, we adjust B17's results to account for the slightly boosted $^{56}$Ni production found by S17, so that $1.15 \msol$ explosions yield light curves with $\dmb= \unit[0.7]{mag}$.  Above $1.15 \msol$, we extend the mapping with an ad hoc linear relation between WD mass and $\dmb$.  Finally, in between $0.85$ and $1.15 \msol$, we  roughly convolve B17's non-LTE radiation transport results with S17's nucleosynthesis.  This leads to the mapping shown in Figure \ref{fig:dm15map}.

We now turn to a theoretical prediction for the evolution of the exploding WD mass using the \texttt{SeBa} binary population synthesis code \citep{port96a,toon12a}.  We employ \texttt{SeBa} to simulate a large number of binaries focusing on those that lead to a merger between two WDs. The simulations include stellar evolution and interactions such as mass transfer and accretion, angular momentum loss, and gravitational wave emission. 

We only consider double WD progenitors that explode promptly as sub-$ \mch$ detonations, before they can evolve into super-$\mch$  remnants.  We are agnostic as to the exact  explosion mechanism, as long as it occurs shortly after the onset of mass transfer and in such a way that the light curve of the SN~Ia is primarily determined by $M_1$, the mass of the more massive WD, which we constrain to be a C/O WD.  Explosion mechanisms that fit these criteria can occur in merging double WD systems via ``dynamically-driven double degenerate double detonations'' \citep{guil10} or direct carbon ignitions \citep{pakm10}.  Stably mass-transferring double WD systems may also lead to double detonation SNe~Ia \citep{bild07}, but recent work suggests that even extreme mass ratio double WD systems will merge unstably \citep{shen15a,brow16b}, so we continue under this assumption for simplicity.

The \texttt{SeBa} simulations used here are based on the primary $\alpha \gamma$-Abt model in \cite{toon17a}.  In this model, the common envelope (CE) prescription is tuned to best reproduce the observed double WD population  \citep{nele00a,toon12a}.  The $\gamma$-CE prescription \citep{nele00a} is applied with $\gamma=1.75$, unless the binary contains a compact object or the CE is triggered by a tidal instability. In the latter  case, the classical $\alpha$-CE prescription is applied \citep{pacz76, webb84}, with $\alpha\lambda=2$. The initial orbital separations follow a power-law distribution with an exponent of $-1$ \citep{abt83a}. For further information, see \cite{toon17a} and references therein. Note that while we show results using the $\gamma$-formalism in this Letter, the trends remain if we exclusively use the $\alpha$-prescription with $\alpha\lambda=2$.

The retention efficiency of helium has been updated with respect to \cite{toon17a}. Based on recent modeling of helium accretion onto WDs \citep{pier14a, broo16a}, we assume that WDs accrete helium conservatively when the logarithm of the mass transfer rate is between
\begin{eqnarray}
	\log_{10} \left( \frac{ \dot{M}_{\rm upper}}{M_\odot / {\rm yr}} \right) &=& -7.226 + 2.504 \left( \frac{ M_{\rm WD}}{M_\odot} \right) \nonumber \\
	&& -0.805 \left( \frac{ M_{\rm WD}}{M_\odot} \right)^2
\end{eqnarray}
and 
\begin{eqnarray}
	\log_{10} \left( \frac{ \dot{M}_{\rm lower}}{M_\odot / {\rm yr}} \right) &=& -8.918+ 4.099 \left( \frac{ M_{\rm WD}}{M_\odot} \right) \nonumber \\
 	&&-1.232 \left( \frac{ M_{\rm WD}}{M_\odot} \right)^2 ,
\end{eqnarray}
where $M_{\rm WD}$ is the mass of the accreting WD. Outside of this regime, the accretion is assumed to be completely non-conservative. The updated helium retention efficiency leads to less WD mass growth compared to previous assumptions \citep{kato99a, bour13a,ruit13a}.

\begin{figure}
  \centering
  \includegraphics[width=\columnwidth]{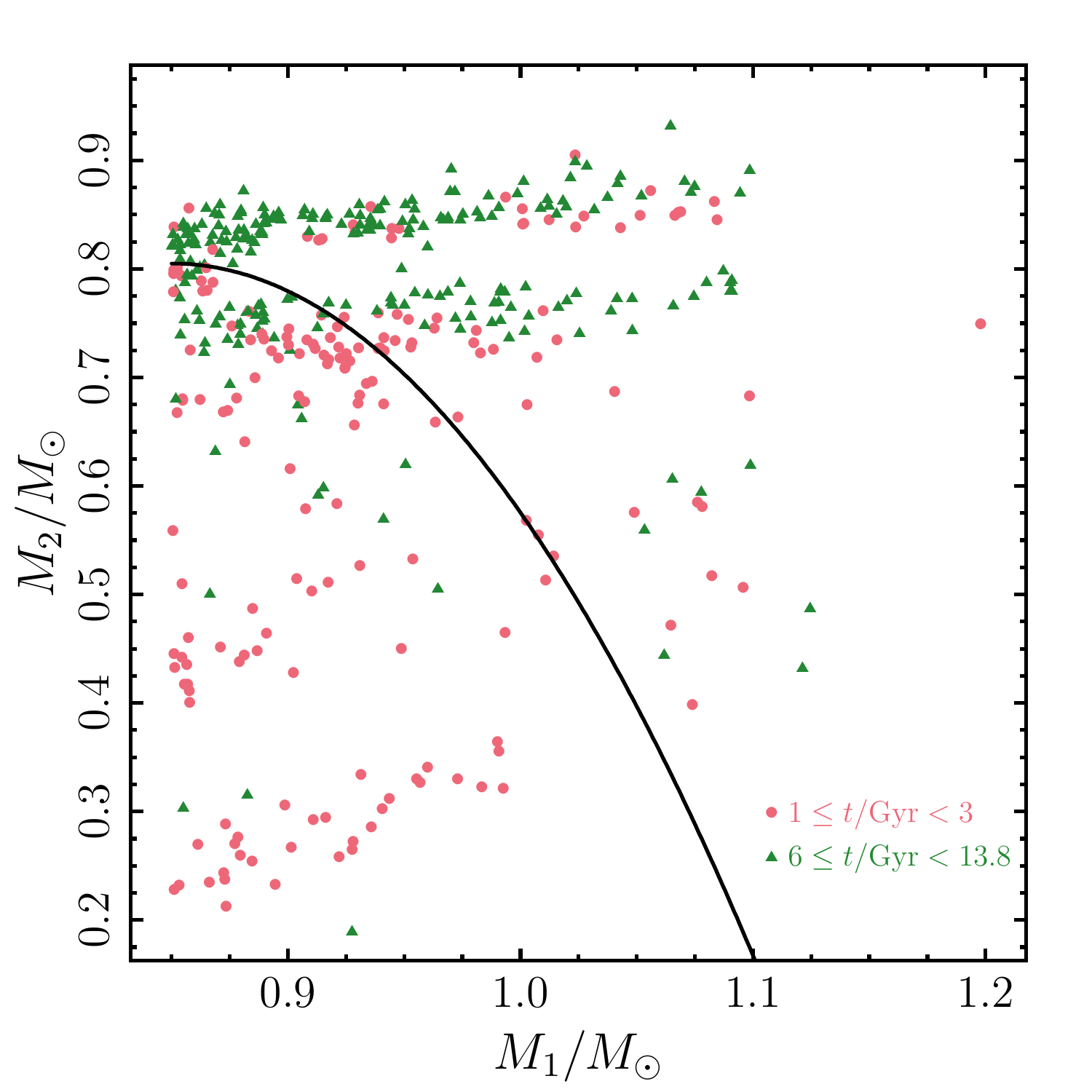}
  \caption{Primary and secondary WD masses at merger for short ($ \unit[1-3]{Gyr}$; \emph{red circles}) and long ($ \unit[6-14]{Gyr}$; \emph{green triangles}) delay times.  We assume binaries above the solid line explode as SNe~Ia.}
  \label{fig:m1m2}
\end{figure}

Figure \ref{fig:m1m2} shows the primary and secondary WD masses at the time of merger for short and long  delay times.  It is clear  that there is an overabundance of $\sim 0.875 \msol + 0.825 \msol$ mergers in the old population compared to the young population.  These primary masses are what we assume lead to SN 1991bg-like SNe; thus, if the currently theoretically uncertain criterion for which mergers lead to subluminous SNe includes only these binaries with relatively massive secondaries, the theoretical $ \dmb$ distribution will shift toward subluminous SNe in older populations.

So as to maximize SN 1991bg-likes in old populations while including as many SNe~Ia overall as possible, we impose a quadratic minimum secondary mass as shown by the solid line in Figure \ref{fig:m1m2}.  While ad hoc, there is a physical basis for our chosen criterion.  More massive secondaries yield more directly impacting accretion streams, and more massive primaries have higher gravitational potentials.  Both of these effects lead to higher temperature hotspots during the merger, which more easily initiate detonations, suggesting a minimum secondary mass that varies inversely with primary mass. We note that the often-used $M_1+M_2 > \mch$ constraint does not reproduce the observed luminosity function evolution; such a constraint yields too many subluminous SNe~Ia in young stellar populations.

\begin{figure}
  \centering
  \includegraphics[width=\columnwidth]{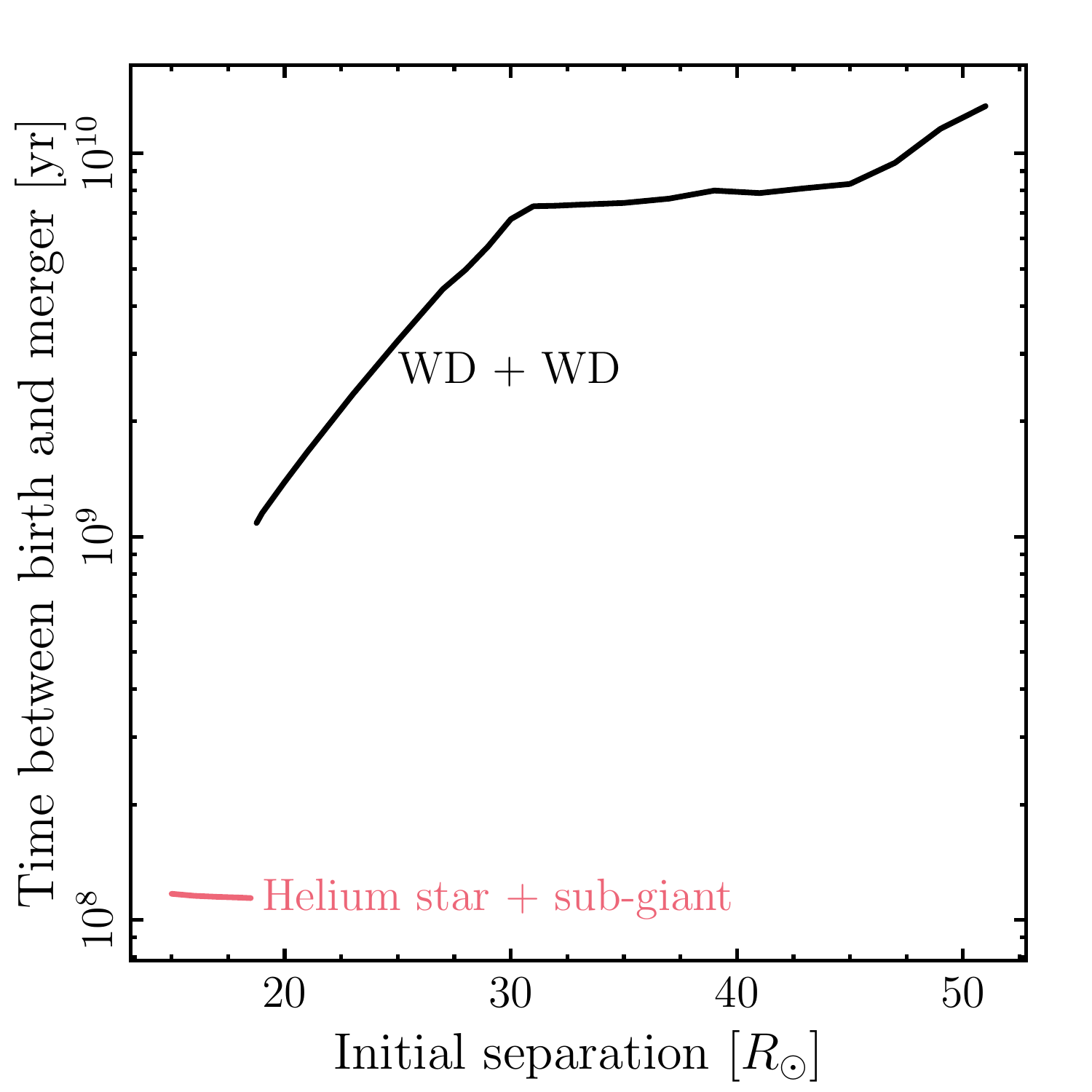}
  \caption{Time between birth and merger vs.\ initial separation for $5.5 \msol +3.5 \msol$ binaries.  Separations that lead to helium star -- sub-giant mergers are shown in red; separations that yield double WD mergers are shown in black.}
  \label{fig:tvsa}
\end{figure}

In order to understand the relative overproduction of WD binaries with masses $\sim 0.875 \msol + 0.825 \msol$ in the older population, we consider the evolution of main sequence binaries with masses $5.5 \msol +3.5 \msol$, which are the main progenitors of these double WD systems.  Figure \ref{fig:tvsa} shows the time between the birth of a $5.5 \msol +3.5 \msol$ binary and the merger of its two components vs.\ initial separation.  For initial separations $< 19 \, R_\odot$, the secondary star fills its Roche lobe as it crosses the Hertzsprung gap before the primary becomes a WD, resulting in a helium star -- sub-giant merger.  For wider initial separations, this mass transfer occurs later, when the primary is already a WD, and leads to a common envelope and a surviving double WD binary   whose separation and gravitational inspiral time are correlated with the initial separation.  Such systems with merger times $\unit[1-3]{Gyr}$ do exist and will lead to subluminous SNe~Ia in young populations, but they are significantly outnumbered by those with merger times $ \unit[6-14]{Gyr}$; thus, we find more faint SNe in old stellar populations.

\begin{figure}
  \centering
  \includegraphics[width=\columnwidth]{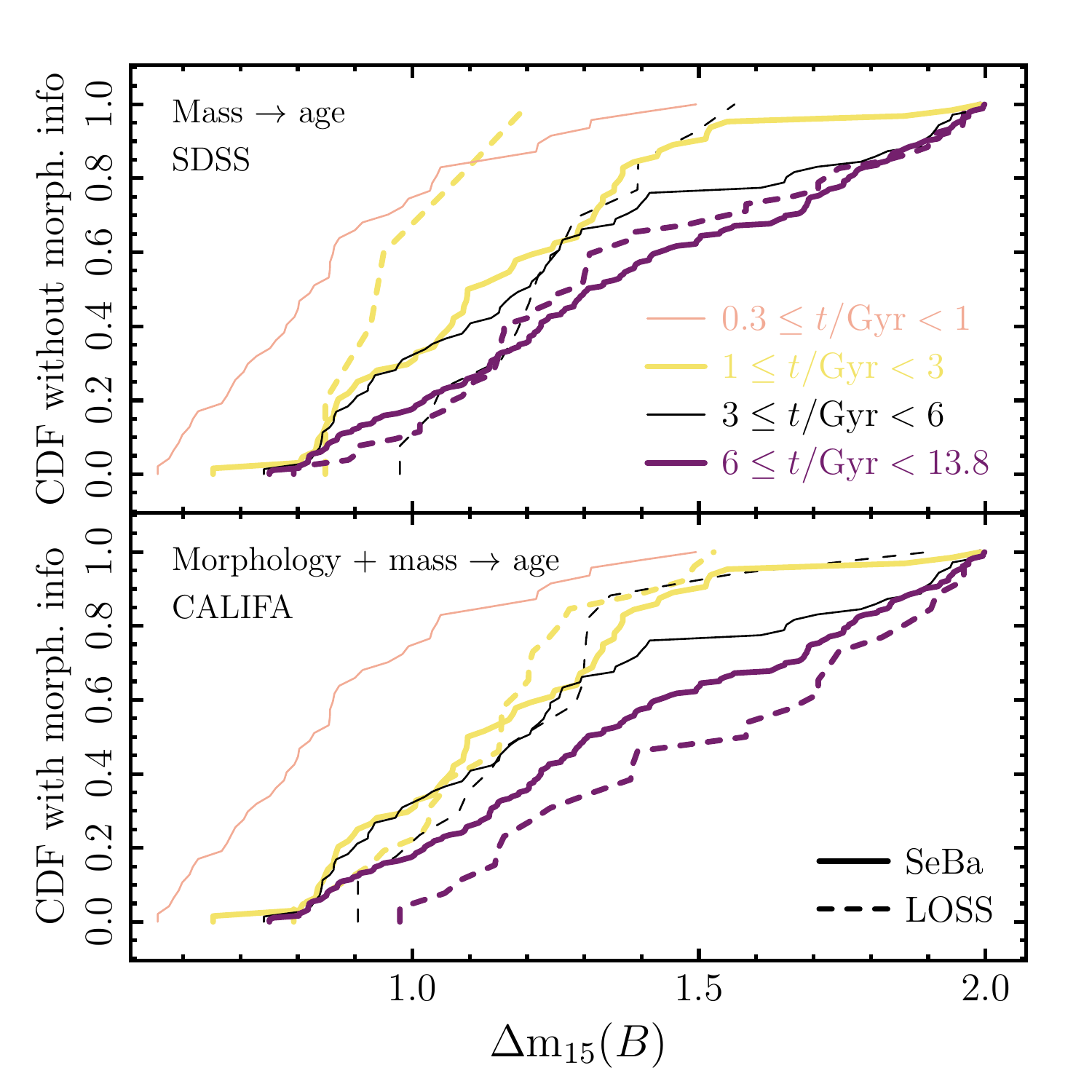}
  \caption{Cumulative distribution functions of $\dmb $ from the LOSS data (\emph{dashed lines}, \S \ref{sec:loss}) for different age bins as labeled, compared to  \texttt{SeBa} CDFs (\emph{solid lines}, \S \ref{sec:seba}).  The LOSS CDFs in the top panel use relations derived from SDSS data to estimate ages from  galaxy masses; stellar ages in the bottom panel are inferred from galaxy masses and morphologies using data from the CALIFA survey.  The youngest age bin's theoretical CDF does not have an observational counterpart.  (The data used to create the observational CDFs in this figure are available in the online journal.)}
  \label{fig:cdfs}
\end{figure}

The resulting theoretical CDFs for four age bins are shown in Figure \ref{fig:cdfs}.  The CDFs are significantly different from one another and in qualitative agreement with the observed CDFs from LOSS: younger stellar populations host fewer dim SNe~Ia than older populations.  Quantitative discrepancies certainly exist between the theoretical and observed CDFs.  However, given the approximations in our analysis, our goal in this Letter is to merely demonstrate that double WD mergers have the capability to explain the evolution of the SN~Ia luminosity function.  Note that the lack of young, low-luminosity galaxies in the LOSS sample precludes a comparison to the theoretical CDF of the youngest age bin.

The overall SN~Ia rates from our binary population synthesis calculations range from $ 10.0\E{-15} \, M_\odot^{-1} {\rm \, yr^{-1}}$ $ \unit[1-3]{Gyr}$ after birth to $ 7.3\E{-15} \, M_\odot^{-1} {\rm \, yr^{-1}}$ $ \unit[6-14]{Gyr}$ after birth.  These rates are $3-10$ times lower than the observed delay time distribution \citep{maoz17b}.  However, this disagreement is within current uncertainties given the similar factor of a few discrepancy between the observed and theoretical local double WD space density \citep{maoz17a,toon17a}.

% -----------------------------------------------------------
% -----------------------------------------------------------

\section{Conclusions}
\label{sec:conc}

In this Letter, we have shown that prompt detonations in double WD systems can qualitatively explain the time evolution of the SN~Ia luminosity function.  Given the many approximations we have made, precise agreement between theory and observations is not  expected and indeed is not achieved; we simply demonstrate a proof of concept.

The largest observational uncertainties relate to our derivation of stellar ages from global galaxy properties such as mass and morphology.  Future work can improve these age estimates by including information, particularly star formation proxies, local to the SN~Ia site. Furthermore, upcoming surveys such as the Zwicky Transient Facility and the Large Synoptic Survey Telescope will greatly increase the numbers of SNe~Ia, reducing Poisson errors and allowing more finely grained age bins, particularly for the low mass, young galaxies not probed by LOSS.

The theoretical side of this work relies on several assumptions that will be improved in the near future.  A combination of more precise detonation simulations and non-LTE radiative transfer calculations is currently underway and will better quantify the mapping between exploding WD mass and $\dmb$.  Future merger simulations will determine the minimum secondary mass that can trigger the primary WD to explode, obviating the need to impose an ad hoc constraint.  Furthermore, concrete progress is being made in modeling common envelopes, which will reduce one of the largest binary population synthesis uncertainties.

A more quantitative study measuring and reproducing the evolution of the SN~Ia luminosity function awaits these and other improvements.  Our work in this Letter simply demonstrates that prompt detonations in double WD systems have the capacity to match this evolution, a constraint that any progenitor scenario attempting to explain the majority of SNe~Ia must confront.

% -----------------------------------------------------------
% -----------------------------------------------------------

\acknowledgments

We gratefully acknowledge Samaya Nissanke and the organizers of the Physics of Extreme Gravity Stars workshop, where some of this work was carried out.  We thank Alison Miller, Peter Nugent, and Mark Sullivan for helpful discussions and Anna Gallazzi for sharing data.  KJS receives support from the NASA Astrophysics Theory Program (NNX15AB16G and NNX17AG28G).  OG is supported by an NSF Astronomy and Astrophysics Fellowship under award AST-1602595.  ST gratefully acknowledges support from the Netherlands Research Council NWO (grant VENI [\#639.041.645]).

% -----------------------------------------------------------
% -----------------------------------------------------------

\end{document}